\documentstyle[preprint,aps,epsfig]{revtex}
\begin{document}
\draft
\title{Local field distribution near corrugated interfaces: \\
       Green's function formulation}
\author{K. W. Yu and Jones T. K. Wan}
\address{Department of Physics, The Chinese University of Hong Kong,\\
         Shatin, New Territories, Hong Kong, China}
\maketitle

\begin{abstract}
We have developed a Green's function formalism to compute the local field 
distribution near an interface separating two media of different 
dielectric constants.
The Maxwell's equations are converted into 
a surface integral equation; thus it greatly simplifies the solutions 
and yields accurate results for interfaces of arbitrary shape.
The integral equation is solved and the local field distribution is 
obtained for a periodic interface.
\end{abstract}
\vskip 5mm

The investigation of the electric field distribution in heterogeneous 
dielectric media has been an established topic \cite{Jackson}. 
However, most studies have been restricted to interfaces of simple geometry.
These studies are important because the detailed form of the local field 
distribution is responsible for the shifting of electronic states at 
semiconductor interfaces \cite{Hofer}. 
The same form is also important for interpretation of scanning electrostatic 
force microscopy experiments \cite{Binnig,Hu}.
However, real surfaces must have some degree of roughness. 
Such an imperfection not only influences the local field distribution 
but it also determines the implications on physical systems \cite{Rahman}.

Theoretical calculation of the local field distribution requires 
a detailed solution of Maxwell's equations.
It is a formidable task for interfaces of complex geometry because of
the complicated boundary conditions.
Thus various attempts have been made in the literature to avoid the match 
of boundary conditions at each interface of the system \cite{Girard}.
In this paper, we develop a Green's function formalism to compute
the local field distribution for an arbitrary interface separating 
two media of different dielectric constants.
Moreover, the local field distribution is calculated near 
a periodic interface.

The object of this investigation is two-fold. 
Firstly, we aim at developing practical computational methods for 
calculating local field distribution near interfaces of arbitrary shape. 
Secondly, the impact of geometric resonance on local field distribution will
be studied.
The results show that the interface profile will provide an extra 
degree of freedom for tunable resonance enhancement of local field.

{\em Integral equation formalism}.
The electrostatic potential satisfies the Laplace's equation:
\begin{eqnarray}
\nabla \cdot [\epsilon({\bf r}) \nabla \phi({\bf r})] = -4\pi \rho({\bf r}),
\label{electrostatics}
\end{eqnarray}
with standard boundary conditions on the interface, where $\rho({\bf r})$
is the free charge density, $\epsilon({\bf r})$ equals $\epsilon_2$ in the 
host and $\epsilon_1$ in the embedding medium.

In a conference paper \cite{Yu2000}, we developed an integral equation 
approach to compute the Green's function for an interface separating 
two media of different dielectric constants.
By using the Green's identity, we were able to convert the volume integral
equation into a surface integral equation; thus it greatly simplifies
numerical solutions and yields accurate results for interfaces of
arbitrary shape. Here we re-iterate the formalism to establish notations.

Let $V_1$ and $V_2$ be the volume of the embedding and host medium, 
separated by an interface $S$.
Denoting $\theta({\bf r})=1$ if ${\bf r} \in V_1$ and 0 otherwise,
leads to an integral equation \cite{Yu2000}:
\begin{eqnarray}
[1-u\theta({\bf r})]\phi({\bf r})
  =\phi_0({\bf r})  
    + {u\over 4\pi} \oint_S dS'\left[ \hat{\bf n}'
      \cdot \nabla' G({\bf r}, {\bf r}') \right] \phi({\bf r}'),
\label{integral}
\end{eqnarray}
where $u=1-\epsilon_1/\epsilon_2$, $\hat{\bf n}'$ is unit normal to $S$, 
$G({\bf r}, {\bf r}')=1/|{\bf r}-{\bf r}'|$ and 
$\phi_0$ is the solution of $\nabla^2 \phi_0=-4\pi \rho/\epsilon_2$.

Accordingly, our approach aims to solve a surface integral equation for
the potential at the expense of a two-step solution:
\begin{enumerate}
\item step 1: determine $\phi({\bf r})$ for all {\bf r} $\in S$
  by solving Eq.(\ref{integral}), and then
\item step 2: obtain $\phi({\bf r})$ for all {\bf r}
  by using Eq.(\ref{integral}) and the results of step 1.
\end{enumerate}
In step 1, we encounter a singularity when the integration variable
${\bf r}'$ approaches the point of observation {\bf r}.
To circumvent the problem, we take an infinitesimal volume around {\bf r}
and perform the surface integral analytically, we find \cite{Yu2000}
\begin{eqnarray}
\left( 1-{u\over 2}\right) \phi({\bf r})
  =\phi_0({\bf r}) 
    + {u\over 4\pi} \oint'_S
      dS'\left[ \hat{\bf n}' \cdot \nabla' G({\bf r}, {\bf r}') \right]
        \phi({\bf r}'), \ \ \ {\bf r} \in S,
\label{r_in_S}
\end{eqnarray}
where ``prime'' denotes a restricted integration which excludes
${\bf r}'={\bf r}$.
The (surface) integral equation (\ref{r_in_S}) can be solved for
$\phi({\bf r}\in S)$.

{\em Periodic corrugated interfaces}.
Here we apply the integral equation formalism to a periodic interface. 
Suppose the interface profile depends only on $x$, described by $y=f(x)$, 
where $f(x)$ is a periodic function of $x$ with period $L$: $f(x+L)=f(x)$. 
Thus medium 1 occupies the space $y<f(x)$ while medium 2 occupies 
the space $y>f(x)$ separated by the interface at $y=f(x)$.
The external field is ${\bf E}_0$ and 
$\phi_0({\bf r})=-{\bf E}_0 \cdot {\bf r}$.
For a periodic system, $\phi({\bf r})$ is a periodic function of the 
lattice vector ${\bf T}$. In what follows, we adopt similar treatment
as the Korringa, Kohn and Rostoker (KKR) method \cite{KKR}
and rewrite the integral equation as:
\begin{eqnarray}
\left( 1-{u\over 2}\right) \phi({\bf r})
  =-{\bf E}_0 \cdot {\bf r} 
    + {u\over 4\pi} \oint'_S
      dS' \tilde{G}({\bf r}, {\bf r}') \phi({\bf r}'),
\label{unit-cell}
\end{eqnarray}
where the integration is performed within a {\em unit cell}. 
The structure Green's function (Greenian) is given by \cite{KKR}:
\begin{eqnarray}
\tilde{G}({\bf r}, {\bf r}')=\sum_{\bf T} \hat{\bf n}' \cdot \nabla' 
    G({\bf r}, {\bf r}'+{\bf T}) 
=\sum_{m} {2\hat{\bf n}' \cdot ({\bf r}-{\bf r}'-m\hat{\bf x})
  \over |{\bf r}-{\bf r}'-m\hat{\bf x}|^2 }.
\end{eqnarray}
In view of the complicated calculus, the evaluation of the Greenian 
is a formidable task. 
However, by invoking the complex notation $z=x+iy$, $z'=x'+if(x')$, 
we are able to evaluate the Greenian analytically:
\begin{eqnarray}
\tilde{G}(x,y;x',f(x')) = {\rm Re}\ \sum_{m} {2idz'/dx' \over z - z' -m} 
  = {2\pi [f'(x') \sin 2\pi (x - x') - \sinh 2\pi (y - f(x')) ] \over            
    \cos 2\pi (x - x') - \cosh 2\pi (y - f(x')) }.
\label{Greenian}
\end{eqnarray}
Eq.(\ref{Greenian}) is readily derived by using the identity \cite{identity}
for the infinite sum:
$$
\sum_{m=-\infty}^\infty {1\over z-m} = \pi \cot (\pi z),
$$
$z$ being a complex number, and by taking the real part of the result.
Eq.(\ref{Greenian}) is a truly remarkable result -- the analytic expression
is valid for an arbitrary interface profile.
For a planar interface, $f(x)=f'(x)=0$, then $\tilde{G}=0$ and $\phi=0$
on the interface, as expected.
If the point of observation $(x,y)$ is located at the interface, 
the Greenian has a finite limit as $x' \to x$:
\begin{eqnarray}
\tilde{G}(x,f(x);x',f(x')|x'\to x)={f''(x)\over 1+(f'(x))^2}.
\end{eqnarray}
We first solve Eq.(\ref{unit-cell}) for the potential $\phi(x,f(x))$ 
right at the interface:
\begin{eqnarray}
\left( 1-{u\over 2}\right) \phi(x,f(x))
  =-E_0 f(x) 
   + {u\over 4\pi} \int dx' \tilde{G}(x,f(x);x',f(x')) \phi(x',f(x')).
\end{eqnarray}
Then we use Eq.(\ref{integral}) to find the potential at any 
arbitrary point $(x,y) \notin S$, using the potential at the interface. 
\begin{eqnarray}
[1-u\theta(f(x)-y)]\phi(x,y)
  =-E_0 y  
   + {u\over 4\pi} \int dx' \tilde{G}(x,y;x',f(x')) \phi(x',f(x')),
\end{eqnarray}
where $\theta(x)=0$ for $x<0$, and $\theta(x)=1$ for $x>0$.
The electric field is computed by the negative gradient of the potential.
We find the electric field in medium 2:
\begin{eqnarray}
E_x &=& -{\partial \phi(x,y) \over \partial x}
= -{u\over 4\pi} \int dx' 
  {\partial \tilde{G}(x,y;x',f(x')) \over \partial x}
  \phi(x',f(x')), \\
E_y &=& -{\partial \phi(x,y) \over \partial y} 
= E_0 - {u\over 4\pi} \int dx' 
  {\partial \tilde{G}(x,y;x',f(x')) \over \partial y} 
  \phi(x',f(x')),
\end{eqnarray}
and similar expressions in medium 1.
The partial derivatives of the Greenian can be computed analytically from
Eq.(\ref{Greenian}).

{\em Solution by mode expansion}.
To solve the integral equation, we express the potential at an 
arbitrary point into a mode expansion:
\begin{eqnarray}
\phi(x,y) = \sum_{jk} C_{jk} \psi_j(x) \xi_k(y-f(x)),
\label{mode}
\end{eqnarray}
where $\psi_j(x)$ and $\xi_k(y)$ are mode functions.
The potential on the interface suffices:
\begin{eqnarray}
\phi(x,f(x)) = \sum_{jk} C_{jk} \psi_j(x) \xi_k(0) 
 = \sum_{j} A_{j} \psi_j(x),
\label{mode-a}
\end{eqnarray}
where $A_j=\sum_k C_{jk} \xi_k(0)$. Here a few remarks are in order
regarding the choice of the mode functions. In theory, the choice of the
mode function is somewhat arbitrary. In practice, these functions should be 
simple and easy to use. Common choice ranges from extended mode functions 
like the Fourier series expansions to localized mode functions 
like the step and triangular functions.
Substituting the mode expansion Eq.(\ref{mode-a}) into Eq.(\ref{unit-cell}), 
the coefficients $A_i$ satisfy the matrix equation:
\begin{eqnarray}
\left[ \left( 1-{u\over 2}\right) {\sf B} - {u\over 4\pi} {\sf M} \right]
 {\bf A} = -E_0 {\bf V},
\end{eqnarray}
where 
\begin{eqnarray}
B_{ij} &=& \int dx \psi_i(x) \psi_j(x),
\\
M_{ij} &=& \int \int dx dx' \psi_i(x) \tilde{G}(x,f(x);x',f(x')) \psi_j(x'),
\\
V_{i} &=& \int dx \psi_i(x) f(x).
\end{eqnarray}
It should be remarked that the mode functions need not be orthonormal and
the matrix {\sf B} is non-diagonal in general.

{\em Numerical results}.
As an illustration, we adopt the interface profile:
$$
f(x)=-\cos {2\pi x \over L} - 0.1 \sin {4\pi x \over L},
$$
where the sine function is added to upset the reflection symmetry about $x=0$.
We adopt the step functions for the mode expansions:
$$
\theta_i^h(x)=1,\ \ \ x_i-{h\over 2} < x < x_i+{h\over 2}\ \ \ {\rm and}\ \
\theta_i^h(x)=0,\ \ \ {\rm otherwise}, 
$$
where $h$ is the width of the step function. 
In what follows, we adopt 100 step functions, equally spaced
in the unit interval $x \in (-L/2,L/2]$. 
The integrals Eqs.(15)--(17) can be readily performed.

We have attempted the following calculations:
\begin{enumerate}
\item For positive $\epsilon_1/\epsilon_2$ ratio,
we find $\phi(x,f(x))$ right at the interface and compute 
$\delta \phi(x)=\phi(x,f(x))-\phi_0(x,f(x))$.
The planar interface result is $\delta \phi$ = 0.

\item For negative $\epsilon_1/\epsilon_2$ ratio, there exists a surface 
plasmon resonance. We find $\delta \phi(x)$ right at the interface 
near a geometric resonance of the interface. 
\end{enumerate}
The geometric resonance is characterized by nontrivial solution of 
Eq.(14) even in the absence of $E_0$.
Without loss of generality, we choose $E_0=1$.
Fig.1 shows the results for $\delta \phi(x)$ and $E_y$ right at 
the interface. Two parameters of $u=-1,-9$ are used in the calculations.
It is clear that the amplitude of $\delta \phi$ increases with 
the dielectric contrast.  
It is more interesting to compute the potential near a geometric resonance
of the interface. Fig.2 shows the results for $\delta \phi(x)$ at a positive 
value of $u=3.3645$. There is a giant enhancement of the potential. 
We also compute the $y$-component of the electric field at a fixed distance 
$y=1.1$. 
For negative values of the $\epsilon_1/\epsilon_2$ ratio, there is
strong enhancement of the local electric field.
In both cases, $E_y$ exhibits similar behavior as that of $\delta\phi$. 
It should be remarked that $\delta \phi$ and $E_y$ here 
are not computed at the same position.

We have performed similar calculations for the triangular function and 
obtained essentially the same results. However, for the triangular 
functions, the {\sf B} matrix is tridiagonal. 
Alternatively, we repeat the calculation by using the Fourier series 
expansion. 
\begin{eqnarray}
\phi(x,f(x))=a_0 + \sum_{n} \left( a_n \cos {2n\pi x\over L} 
  + b_n \sin {2n\pi x\over L} \right).
\end{eqnarray}
We have adopted $n=1,2,3$ harmonics in the Fourier series expansions. 
For $u=-1,-9$, the results (not shown here) of $\delta\phi(x)$ agree 
with the calculations by using the step function expansions. 
However, the $u=3.3645$ results exhibit some deviations. 
This can be clearly seen from Fig.2 that the local field distribution 
contains rich harmonics near a geometric resonance of the interface.

The Fourier coefficients for a few lowest harmonics are listed in Table I. 
From these coefficients, we reproduced $\delta\phi$ at $x=0$ 
and computed the error. 
As is evident from Table I, the error is small for the $u=-1, -9$ cases,
while there is a large error for the resonant case. 
Thus, to achieve more accurate results, higher harmonics need to be 
included in the resonant case.

In summary, we have developed a Green's function formalism to calculate 
the local field distribution of an interface of arbitrary shape.
The solution is accomplished first by solving the potential at the
interface, followed by determining the potential at any arbitrary
point by performing an integral over the interface.
Thus it greatly simplifies and yields solutions for interfaces
of arbitrary shape.
Moreover, the local field distribution have been calculated near
a periodic interface.

This work was supported by the Research Grants Council of the Hong Kong 
SAR Government under grant CUHK 4284/00P. K. W. Y. acknowledges initial
collaboration with Professor Hong Sun.

\newpage

\noindent Table I: Fourier coefficients for the potential at interface.
\[
\begin{array}{c c c c c c c c c}
\hline \\
  & {\rm constant} & \cos 2\pi x & \cos 4\pi x & \cos 6\pi x 
    & \sin 2\pi x & \sin 4\pi x & \sin 6\pi x & {\rm error} \\
  & a_0 & a_1 & a_2 & a_3 & b_1 & b_2 & b_3 &  \\ \\ \hline \\
{\rm interface\ profile} & 0.0000 & -1.0000 & 0.0000 & 0.0000 
    & 0.0000 & -0.1000 & 0.0000 & 0.0000 \\ \\
u=-1.0 & -0.1739 & -0.3268 & -0.0337 & 0.0018 
    & -0.0083 & -0.0321 & -0.0066 & 0.0004 \\ \\
u=-9.0 & -0.1228 & -0.8065 & -0.0246 & 0.0033 
    & -0.0060 & -0.0797 & -0.0050 & -0.0001 \\ \\
u=3.3645 & 12789 & -6935 & 3822 & -3005 
    & 2392 & 483 & 2676 & -166 \\ \\ \hline
\end{array}
\]

\begin{figure}[h]
\caption{$\delta\phi(x)$ (thin lines) and $E_y$ (thick lines) plotted
against $x$ for two values of dielectric constrasts $u=-1.0$ and $-9.0$.} 
\label{Fig1}
\end{figure}

\begin{figure}[h]
\caption{$\delta\phi(x)$ (solid line) and $E_y$ (dashed line) plotted
against $x$ for $u=3.3645$. There is a giant enhancement of the local field.}
\label{Fig2}
\end{figure}

\newpage
\centerline{\epsfig{file=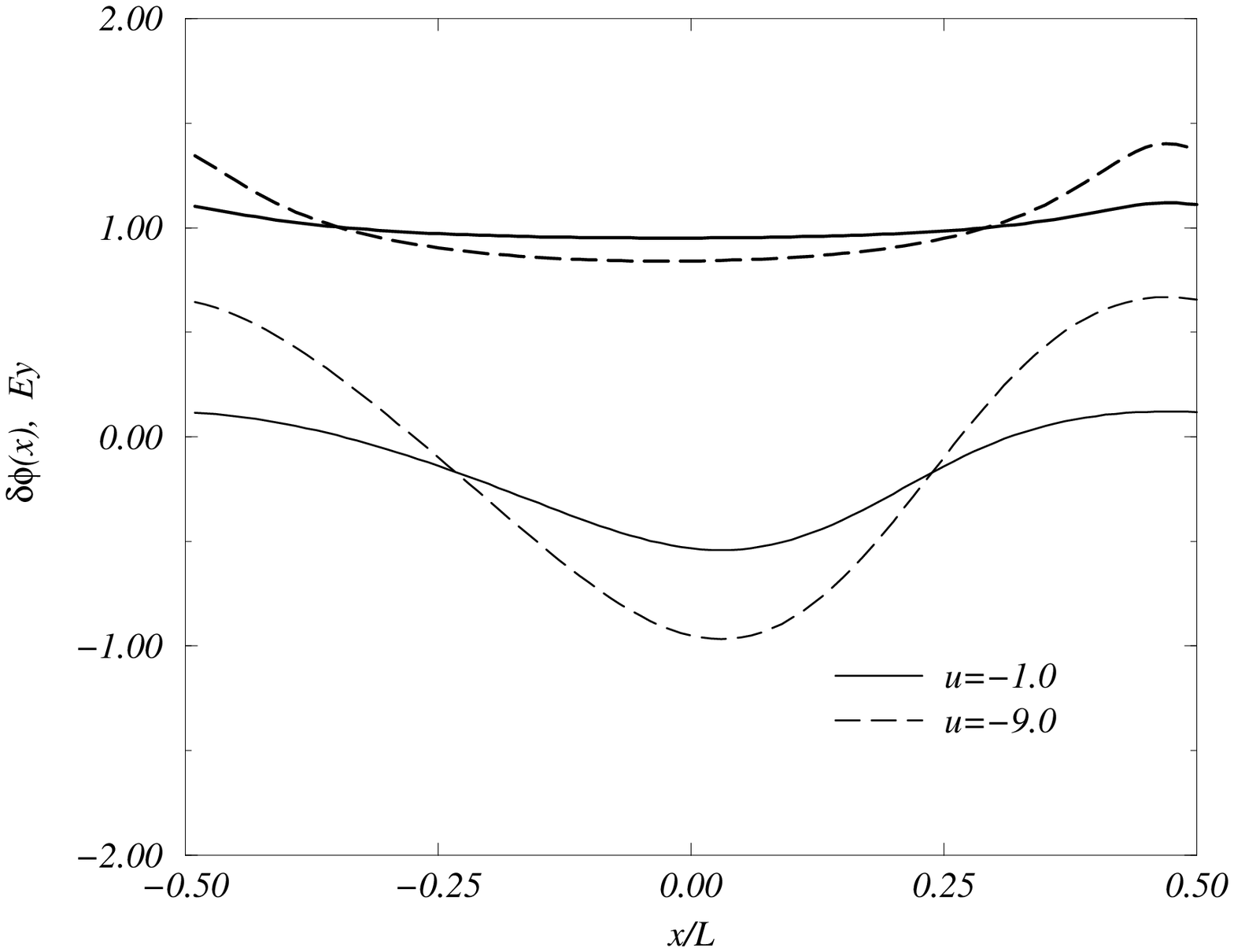,width=\linewidth}}
\centerline{Fig.1/Yu and Wan}

\newpage
\centerline{\epsfig{file=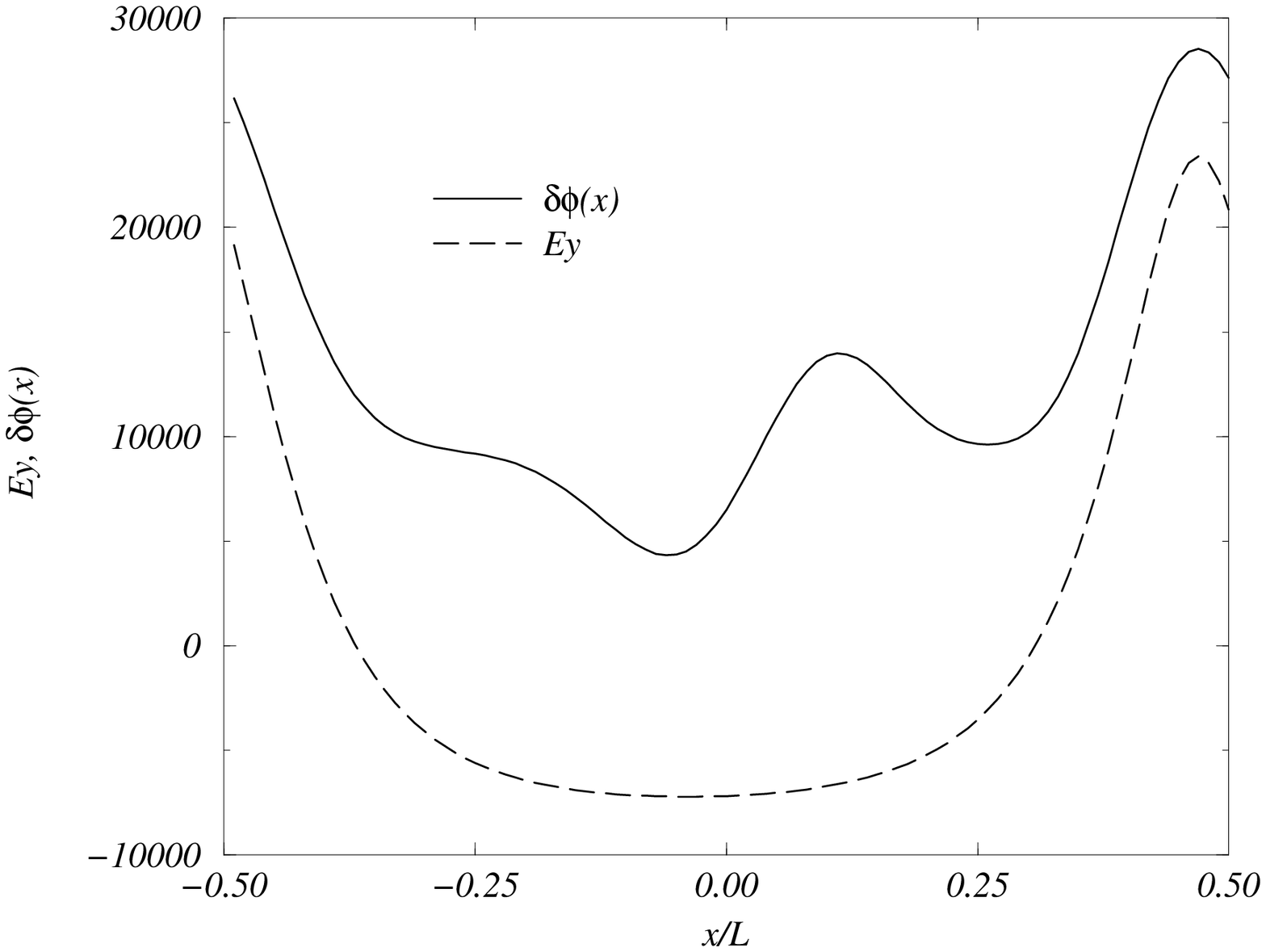,width=\linewidth}}
\centerline{Fig.2/Yu and Wan}

\end{document}